%% file: main.tex
\documentclass[conference]{IEEEtran}
\IEEEoverridecommandlockouts
\usepackage{cite}
\usepackage{amsmath,amssymb,amsfonts}
\usepackage{algorithmic}
\usepackage{graphicx}
\usepackage{textcomp}
\usepackage{xcolor}
\usepackage{colortbl}
\usepackage{multirow}
\usepackage[hyphens]{url}                 %
\usepackage[hidelinks,breaklinks]{hyperref}%
\usepackage{xurl}                         %
\def\BibTeX{{\rm B\kern-.05em{\sc i\kern-.025em b}\kern-.08em
    T\kern-.1667em\lower.7ex\hbox{E}\kern-.125emX}}

\usepackage{booktabs}
\usepackage{tabularx}

\begin{document}

\title{Beyond More Context: How Granularity and Order Drive Code Completion Quality}

\author{\IEEEauthorblockN{Uswat Yusuf}
\IEEEauthorblockA{\textit{REALISE Lab} \\
\textit{Concordia University}\\
Montreal, Canada \\
omosewaeniola@gmail.com}
\and
\IEEEauthorblockN{Genevieve Caumartin}
\IEEEauthorblockA{\textit{REALISE Lab} \\
\textit{Concordia University}\\
Montreal, Canada \\
genevieve.caumartin@mail.concordia.ca}
\and
\IEEEauthorblockN{Diego Elias Costa}
\IEEEauthorblockA{\textit{REALISE Lab} \\
\textit{Concordia University}\\
Montreal, Canada \\
diego.costa@concordia.ca}
}

\maketitle

\begin{abstract}
Context plays an important role in the quality of code completion, as Large Language Models (LLMs) require sufficient and relevant information to assist developers in code generation tasks.  
However, composing a relevant context for code completion poses challenges in large repositories: 
First, the limited context length of LLMs makes it impractical to include all repository files. 
Second, the quality of generated code is highly sensitive to noisy or irrelevant context. 
In this paper, we present our approach for the ASE 2025 Context Collection Challenge.
The challenge entails outperforming JetBrains baselines by designing effective retrieval and context collection strategies. We develop and evaluate a series of experiments that involve retrieval strategies at both the file and chunk levels. We focus our initial experiments on examining the impact of context size and file ordering on LLM performance. Our results show that the amount and order of context can significantly influence the performance of the models. 
We introduce chunk-based retrieval using static analysis, achieving a 6\% improvement over our best file-retrieval strategy and 16\% over the no-context baseline for Python in the initial phase of the competition. 
Our results highlight the importance of retrieval granularity, ordering and hybrid strategies in developing effective context collection pipelines for real-world development scenarios.             
\end{abstract}

\section{Introduction}
Code completion plays a crucial role in modern software development by increasing the productivity of engineers \cite{husein2025large}. Studies have shown that developers see significant benefits in code completion as it reduces code iteration time \cite{tabachnyk2022ml}, thus improving their productivity. 
Recent advances in LLMs have accelerated this trend with models like CodeBERT\cite{feng2020codebertpretrainedmodelprogramming}, StarCoder\cite{li2023starcodersourceyou}, CodeLlama \cite{rozière2024codellamaopenfoundation}, which offer highly contextual code suggestions and generation capabilities. 
These models are trained on large code corpora and often fine-tuned to handle diverse programming tasks across domains. 

The effectiveness of large language models (LLMs) depends on the quality and relevance of the input context. In real-world development scenarios, including all source code files can exceed the model's context window. Moreover, overloading a model's context window can lead to noise or affect its performance\cite{liu2023lost}. As such, the design of effective context collection strategies has become a critical factor in real-world deployment of code completion models.    

The ASE 2025 Context Collection Challenge provided a structured setting to explore this problem. Participants were tasked with developing strategies for the retrieval composition of the context to outperform selected JetBrains baselines. 
The performance of the context collection strategies were evaluated  %
 across two programming languages: Python and Kotlin \cite{ustalov2025contextcollection}. In this paper, we describe our approach to the challenge, outlining the progression of our strategies from file-level to chunk-level retrieval and reflecting on the lessons learned on retrieval granularity using lexical similarity and static analysis. 
 We provide a replication package~\cite{replicationpackage} to support the reproducibility of our experiments.

\section{Experiment Setup}
\subsection{Datasets}
We conduct our experiments on the datasets released as part of the JetBrains Code Completion Competition \cite{jetbrains2025starter}. The competition provided two categories of data: \textit{practice data} for the development and debugging of context retrieval strategies and \textit{public data} for the main evaluation phase. The Python portion of the dataset consists of 47 repositories for the practice split and 247 repositories for the public split, while the Kotlin version consists of 30 repositories for the practice split and 400 repositories for the public split. These splits vary in size and structure, with the public split consisting of more complex repositories and challenging instances \cite{jetbrains-cccc-evalai-2516}. 

Beyond repository count, we also analyze the distribution of code files in each repository to better understand the complexity of the data. 
The practice datasets are generally smaller and have more uniform characteristics (e.g., repository size), while the public data consists of repositories with higher variation and a large number of files.
Table \ref{tab:dataset-stats} summarizes key statistics, including the average and median number of files per repository. These characteristics highlight the complexity of the task at hand, as the context-retrieval strategies must generalize across languages and also across projects of widely varying scale. 

\begin{table}[t]
\centering
\caption{Summary statistics of the JetBrains competition datasets.}
\label{tab:dataset-stats}
\begin{tabular}{lrrrrr}
\toprule
\textbf{Dataset} & {\textbf{\#Repos}} & \multicolumn{4}{c}{\textbf{Files per repository (count)}} \\
\cmidrule(lr){3-6}
& & \textbf{Min} & \textbf{Median} & \textbf{Mean} & \textbf{Max} \\
\midrule
Python (Practice) & 47  & 11   & 130   & 214.3  & 650  \\
Kotlin (Practice) & 30  & 35   & 267    & 289.4   & 557  \\
Python (Public)   & 247 & 4    & 137   & 192.3  & 940 \\
Kotlin (Public)   & 400 & 4   & 254   & 668.8  & 7493 \\
\bottomrule
\end{tabular}
\end{table}

\subsection{Models}
In this study, we evaluate our strategies on the models provided in the JetBrains Competition, including Mellum:4b, Qwen2.5-Coder:7b and Codestral \cite{evalai2025contextcomposition}.
\textbf{Mellum:4b} is a code-specific model developed by JetBrains that is fine-tuned on real-world code completion scenarios. It supports a range of programming languages and has a context window of 8,192 tokens \cite{semenkin2025mellum}. 
\textbf{Qwen2.5-Coder:7b} was introduced by Alibaba Cloud and the model is part of the Qwen2.5-Coder series. It is built on the Qwen2.5 architecture and fine-tuned on code-based tasks, achieving strong results across benchmarks\cite{hui2024qwen25codertechnicalreport}. In the competition setting, the context window was limited to 16,000 tokens. 
\textbf{Codestral:22B} is a code model developed by Mistral AI designed with fill-in-the-middle capabilities for code completion~\cite{mistral2024codestral}. Like Qwen2.5-Coder, it has a 16,000 token context window in the competition environment.

\subsection{Code Completion Task}
We evaluate the effectiveness of repository-level code completion across multiple large language models (LLMs). 
The task involves the models generating the missing code segment in a \textit{target file} using both the surrounding code and additional code snippets from the repository. The problem is formulated as a Fill-in-the-middle (FIM) task, where the input consists of: 1) \textbf{Prefix: }the code preceding the missing segment, 2)
\textbf{Suffix: }the code following the missing segment, and 3) \textbf{Additional retrieved context: }relevant code snippets from the repository.

The models' objective is to generate a suitable fit for the missing code that bridges the prefix and suffix, leveraging both local and retrieved repository context. 

\subsection{Evaluation Metrics}
JetBrains evaluated the efficiency of the strategy using the \textbf{ average chrF} (character n-gram F-score) \cite{evalai2025contextcomposition} across the three models. 
The chrF score evaluates the overlap between the generated completion and the ground truth sequences based on contiguous sequences of characters, rather than whole words \cite{paul2024benchmarks}, which makes it sensitive to crucial differences in code syntax \cite{evalai2025contextcomposition}.  

\subsection{Baselines}
To show the need for effective context collection strategies to improve the performance of models in code completion, we define a \textit{no-context} baseline alongside the JetBrains \textit{recent} baseline (which was only computed for Python in the Practice phase) to compare the results of our retrieval strategies. 

The \textbf{no-context baseline} involves providing the LLMs with only the prefix and suffix surrounding the masked code without any additional repository context. 

The \textbf{recent baseline} involves selecting a \textit{single} context file from a list of recently \textit{modified} files in the same commit as the missing code that are above ten lines of code; if no qualifying files are found, a random file is selected. Together, these baselines measure the performance of models with no contextual information and when given minimal, randomly selected context as shown in Table \ref{strategy-results(practice)}   

\section{Searching the Best Approach}

\begin{table*}
    \centering
    \caption{Preliminary results of attempted strategies (Practice Phase). We highlight in gray the best results per experiment, and in bold the best overall results.
    }
    \label{strategy-results(practice)}
    \input{tables/practice-results}
\end{table*}

\begin{table*}
    \caption{Results of the best strategies (Public and Private Phase) }
    \label{Public-phase results}
    \input{tables/public-results}

\end{table*}

\subsection{The Impact of the Top-K BM25 Files}

Our first experiment examines the effect of providing top-k BM25 ranked source code files as contextual input to the models. BM25 (Best Match 25) is a lexical ranking algorithm in information retrieval \cite{10381286} based on matching queries to candidate files. As prior work suggests that increasing the context provided to the models can improve accuracy, we measure the effect of increasing the number of source code files ($k$) on the evaluation metric. 

\begin{figure}[h]
    \centering
    \includegraphics[width=1.0\linewidth]{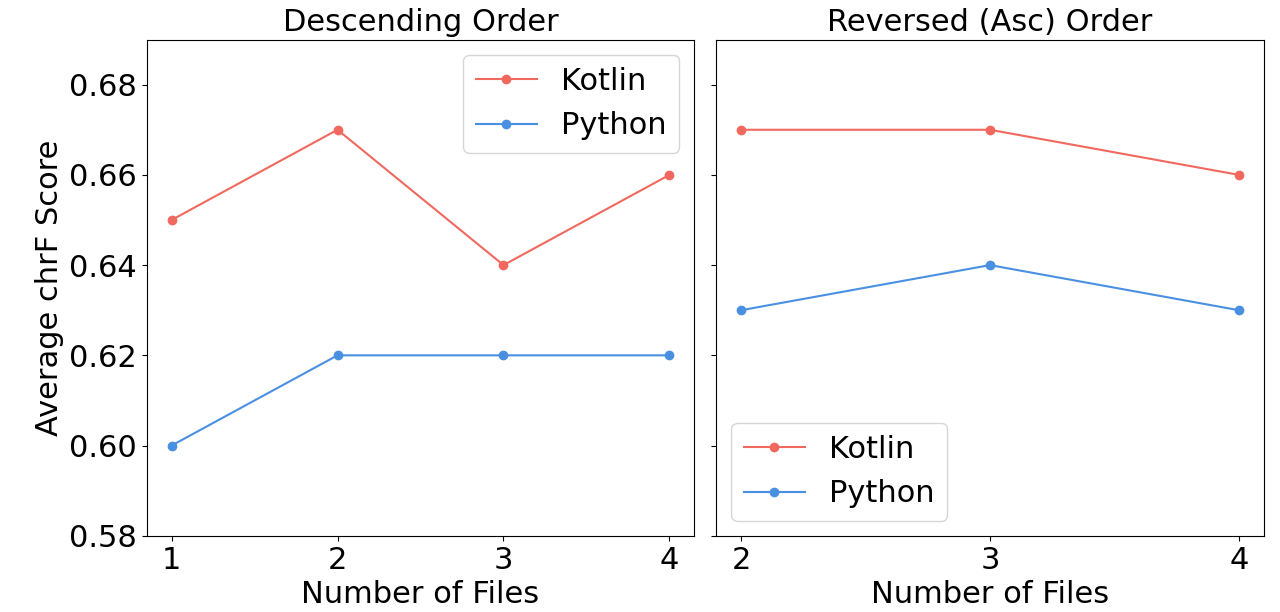} 
    \caption{Number of files vs chrF score} %
    \label{fig:top-files-vs-chrF-score}
\end{figure}

\textbf{Results:} As shown in Figure \ref{fig:top-files-vs-chrF-score} (Descending Order), increasing the number of retrieved files improves performance for both Kotlin and Python. Kotlin achieves its best score at $k=2$ with a $3\%$ improvement over the baseline, while Python also peaks at $k=2$ with a $9\%$ improvement over the baseline. This shows that providing relevant contextual information to LLMs improves their code completion ability. 

\subsection{The Impact of File Ordering}

In this experiment, we evaluate the importance of file ordering on model performance. In our previous experiment, the \textit{top-k files} were ranked using BM25 similarity to the query (prefix and suffix) and arranged in descending order, placing the most relevant files at the beginning of the prompt. To test for ordering bias,  we reversed the order, returning code files in ascending similarity.  This design allowed us to test for potential \textit{primacy or recency bias} in LLMs \cite{liu2023lost} and also account for a practical constraint: when the source code files provided as context exceeds the context window of the models, the JetBrains environment truncates from the left, potentially discarding earlier files \cite{evalai2025contextcomposition}.  

\textbf{Results:} 
As shown in Figure \ref{fig:top-files-vs-chrF-score} (Reversed (asc) Order), reversing the order
of files sometimes yielded small but measurable gains with the best result when $k=3$ which could support the primacy/ recency bias and the truncation
setting. For Python, reversing the order yields a $3\%$ improvement over the default descending order, whereas, for Kotlin, there was no measurable change.

\subsection{The Impact of Chunking the Files.}

In this experiment, we explore chunking files using the individual Treesitter library for both languages~\cite{tree-sitter}, as current research shows that chunks within code syntactic boundaries are better than \textit{syntax-agnostic} chunks \cite{zhang2025cast}. Our initial approach, \textit{standard} chunking, ignores import statements and segments code files into objects, classes, standalone functions, expression statements and docstrings. This design preserves meaningful context in each chunk while reducing noise from import statements which is often of little benefit to the LLM. Our second approach, \textit{method-level chunking}, further splits classes into individual methods, producing smaller segments and allowing the model to access more context. After generating these chunks, we rank them using the BM25 algorithm. 

\textbf{Results:} The chunk-level retrieval strategy outperforms the full file-level retrieval strategy.  In the Practice Phase, the top-5 BM25-ranked chunks yielded the best results for Python with a $3\%$ improvement which motivated us to apply this strategy on Kotlin where it performs comparably to the best file-level approach. Method-level chunking performs worse than standard chunking for Python but achieves similar results for Kotlin. These findings indicate that finer-grained context improves the model's prediction.

\subsection{Impact of Local Scope Context}
As part of an ablation study, we evaluate the effect of trimming the prefix and suffix provided to the model for each code completion instance. For the prefix, we restrict it to the nearest enclosing block of code around the missing code. In Python, this typically corresponded to a class definition, function definition or decorated function and in Kotlin, to an object declaration, class declaration or function declaration. The same logic was applied to the suffix where the trailing portion of the nearest enclosing block was preserved. This experiment tests whether limiting context to the local scope improves completion quality. 

\textbf{Results:} Trimming the prefix and suffix improves performance for both Python and Kotlin as shown in Table \ref{strategy-results(practice)}. In Kotlin, this approach achieves a $3\%$ improvement, while in Python, it reaches an average chrF score of $0.65$ comparable to the best chunk-level retrieval strategy. This shows that implementing this approach can improve code completion quality.

\subsection{Public and Private Phase Evaluations}
In the public phase, as shown in Table \ref{Public-phase results}, the chunk-based retrieval consistently outperforms the file-based retrieval strategy. Our top-5 chunk (reversed order) strategy achieved a chrF score of 0.56 for Python and 0.65 for Kotlin improving over the recent baseline by $8\%$ for Python and $3\%$ for Kotlin. This confirms that fine-grained retrieval remains effective across diverse datasets, as such, it was the basis for our final submission.  %

\noindent
\textbf{Best Python Strategy} 
Based on public phase performance, we select a strategy that retrieves the top-5 BM25-ranked chunks and applies local-scope trimming to the prefix and the suffix. In the private phase this approach achieves an average chrF score of $0.64$, securing third place overall.

\noindent
\textbf{Best Kotlin Strategy}
For Kotlin, our winning strategy also retrieves the top-5 BM25-ranked chunks, but when the combined token count of these chunks is below 2,000, we include three additional chunks to balance efficiency. We also apply local-scope trimming to reduce noise. This achieves an average chrF score of $0.66$ which also secured a tie in the third-place in the contest.

Our results show that chunk-retrieval and local-scope selection scales across diverse datasets and delivers competitive performance across the selected programming languages.

\section{Discussion}

\noindent
\textbf{1. Granularity matters: }
Our final solution's main framework was based on selecting code snippets rather than complete files to give room for more relevant contexts to be sent to the models. After running a number of experiments, we realized that granularity, when done effectively, can improve the performance of models in code completion. 

\noindent
\textbf{2. Ordering Effects Influence LLM Behavior: }
Reversing the order of the selected code snippets yields a small improvement; 
however, an open problem remains whether there would be a notable impact if the context fits within the model's token window. 

\noindent
\textbf{3. Finding a balance between quantity and quality: }
Choosing the right context budget is nontrivial: more context can help, but too much adds noise. Our fixed number of chunks results in repository-dependent token lengths, which can be limiting; tailoring context length to the completion type would likely perform better. 

\noindent
\textbf{4. Beyond Lexical similarity: } 
Our experiments rely mainly on lexical similarity, which is often effective. Recent embedding models enable strong semantic retrieval, and prior work shows hybrid lexical–semantic methods can further improve retrieval quality and model performance. The remaining challenge is selecting models that integrate well with production environments.

\noindent
\textbf{5. Generalization of strategies:}
Strategies affected the languages differently: Kotlin generally outperformed Python, and some methods (e.g., method-level chunking) benefited Kotlin more. Our final solution therefore uses per-language settings—method-level chunking for Kotlin and standard chunking for Python. Cross-language generalization remains an open challenge.
\section*{Acknowledgments}
\emph{Contributions—}  Uswat Yusuf: implementation, methodology, writing; Genevieve Caumartin: mentoring and solution design; Diego Elias Costa: conceptualization, and supervision.
\bibliographystyle{IEEEtran}
\bibliography{references}

\end{document}

%% file: tables/practice-results.tex
\newcommand{\winner}{\cellcolor{gray!15}}

\begin{tabular}{l l|r r|r r|r r|r r}
         \toprule
        & & \multicolumn{2}{c}{\textbf{Average}}
        & \multicolumn{2}{c}{\textbf{Mellum}}
        & \multicolumn{2}{c}{\textbf{Codestral}}
        & \multicolumn{2}{c}{\textbf{Qwen}}
        
        \\
         
        \textbf{Experiment} & \textbf{Strategy}  & \textbf{Python} & \textbf{Kotlin } & \textbf{Python } & \textbf{Kotlin } & \textbf{Python } & \textbf{Kotlin } & \textbf{Python } & \textbf{Kotlin }\\
        \midrule
        Baseline & No-context 
        & 0.57 & 0.65 %
        & 0.56 & 0.64 %
        & 0.63 & 0.70 %
        & 0.51 & 0.61 %
        \\
         & Recent
        & 0.57 & - %
        & 0.56 & - %
        & 0.60 & - %
        & 0.53 & - %
        \\

        \midrule
        \multirow{4}{*}{Top-K Files} & Top 1 file 
        & 0.60 & 0.65 %
        & 0.60 & 0.67 %
        & 0.65 & 0.68 %
        & 0.55 & 0.61 %
        \\
        & Top 2 files 
        & \winner 0.62 & \winner 0.67 %
        & \winner 0.62 & \winner 0.68 %
        & 0.67 & \winner 0.72 %
        & 0.55 & 0.60 %
        \\
        & Top 3 files 
        & 0.62 & 0.64 %
        & 0.61 & 0.62 %
        & \winner 0.68 & 0.70 %
        & \winner 0.56 & 0.61 %
        \\
        & Top 4 files 
        & 0.62 & 0.66 %
        & 0.62 & 0.60 %
        & 0.67 & 0.71 %
        & \winner 0.56 & \winner 0.66 %
        \\

        \midrule
        \multirow{3}{*}{File Ordering} & Top 2 files (reversed order) 
        & 0.63 & 0.67 %
        & 0.63 & \winner 0.70 %
        & 0.68 & 0.69 %
        & \winner 0.59 & 0.64 %
        \\
        & Top 3 files (reversed order)
        & \winner 0.64 & \winner 0.67 %
        & \winner 0.68 & 0.67 %
        & \winner 0.69 & 0.67 %
        & 0.54 & \winner 0.65 %
        \\
        & Top 4 files (reversed order) 
        & 0.63 & 0.66 %
        & 0.64 & 0.66 %
        & \winner 0.69 & \winner 0.70 %
        & 0.57 & 0.64 %
        \\

        \midrule
        
        \multirow{7}{*}{Chunking Files}
        & Top 3 chunks
        & 0.65 & -- %
        & 0.68 & -- %
        & 0.67 & --  %
        & 0.61 & --  %
        \\

        & Top 5 chunks
        & \winner \textbf{0.66} & \winner 0.67 %
        & \winner 0.69 & 0.67 %
        & \winner 0.69 & \winner 0.70  %
        & \winner 0.62 &0.64  %
        \\

        & Top 10 chunks
        & 0.63 & -- %
        & 0.62 & -- %
        & \winner 0.69 & -- %
        & 0.58 & --  %
        \\
        
        & Top 15 chunks
        & 0.62 & -- %
        & 0.59 & -- %
        & 0.68 & -- %
        & 0.58 & -- %
        \\
        
        & Top 20 chunks 
        & 0.61 & -- %
        & 0.57 & -- %
        & 0.66 & -- %
        & 0.59 & -- %
        \\

        & Top 5 method-lvl chunks
        & 0.61 & \winner 0.67  %
        & 0.63 & \winner 0.68  %
        & 0.64 & 0.69 %
        & 0.56 & 0.64 %
        \\
        
        & Top 10 method-lvl chunks
        & 0.62 & 0.66%
        & 0.62 & 0.64  %
        & 0.67 & 0.68 %
        & 0.56 & \winner 0.67  %
             \\
              \midrule
        \multirow{1}{*}{Local-Scope Context} & Top 5 chunks 
        & 0.65 & \textbf{0.69} %
        & 0.68 & \winner 0.68 %
        & \winner 0.71 & \winner 0.71 %
        &  0.59 & \winner 0.67 %
        \\
             \midrule

    \end{tabular}

%% file: tables/public-results.tex
\newcommand{\winner}{\cellcolor{gray!15}}

\begin{tabular}{l l|r r|r r|r r|r r}
         \toprule
        & & \multicolumn{2}{c}{\textbf{Average}}
        & \multicolumn{2}{c}{\textbf{Mellum}}
        & \multicolumn{2}{c}{\textbf{Codestral}}
        & \multicolumn{2}{c}{\textbf{Qwen}}
        
        \\
         
        \textbf{Evaluation} & \textbf{Strategy}  & \textbf{Python} & \textbf{Kotlin } & \textbf{Python } & \textbf{Kotlin } & \textbf{Python } & \textbf{Kotlin } & \textbf{Python } & \textbf{Kotlin }\\
        \midrule
         Baseline & Recent
        & 0.52 & 0.64 %
        & 0.49 & 0.62 %
        & 0.56 & 0.67 %
        & 0.50 & 0.62 %
        \\
        \midrule
        \multirow{3}{*}{Public Phase} 
        & Top 1 file 
        & 0.52 & 0.65 %
        & 0.49 & \winner 0.64 %
        & 0.56 & 0.69 %
        & 0.52 & 0.62 %
        \\
        
        & Top 5 chunks (reversed order) 
        & \winner \textbf{0.56} & 0.65 %
        & \winner 0.52 & 0.62 %
        & \winner 0.60 & \winner 0.71 %
        & \winner 0.55 & \winner 0.63 %
        \\
        & Top 5 method-lvl chunks (reversed order) 
        & 0.54 & 0.65 %
        & 0.51 & \winner 0.64 %
        & 0.57 & 0.69 %
        & 0.53 & 0.62 %
        \\
        & Top 5 chunks (reversed order)(local-scope) 
        & \winner \textbf{0.56} &  \textbf{} %
        & \winner 0.52 &  %
        & \winner 0.60 &   %
        & \winner 0.55 &  %
        \\
        & Top 10 method chunks (reversed)(local-scope) 
        & \textbf{} & \winner \textbf{0.66} %
        &  &0.63  %
        &   & \winner 0.71  %
        &  & \winner 0.64 %
        \\
        \midrule

        \multirow{2}{*}{Private Phase} &
        Best Python strategy 
        &0.64  &  %
        &0.61  &  %
        &0.71  &  %
        &0.61  &  %
        \\

        &
        Best Kotlin strategy
        &  & 0.66 %
        &  & 0.65 %
        &  & 0.69 %
        &  & 0.64 %
        \\

\end{tabular}